\documentstyle[psfig]{mn}

\title[Redshift-space Distortions in the 2dF QSO Power Spectrum]{The 2dF QSO Redshift Survey - VI. Measuring $\Lambda$ and $\beta$ from Redshift-space Distortions in the Power Spectrum}
\author[P.~J.~Outram et al.]
{P.~J.~Outram$^1$, Fiona~Hoyle$^{1,2}$, T.~Shanks$^1$, B.~J.~Boyle$^3$, S.~M.~Croom$^3$,
\newauthor  N.~S.~Loaring$^4$, L.~Miller$^4$, R.~J.~Smith$^{5,6}$\\
$^1$ Dept. of Physics, University of Durham, South Road, Durham DH1 3LE, UK.\\
$^2$ Dept. of Physics, Drexel University, 3141 Chestnut Street, Philadelphia, PA 19104, USA. \\ 
$^3$ Anglo-Australian Observatory, PO Box 296, Epping, NSW 2121, Australia.\\
$^4$ Dept. of Physics, University of Oxford, Nuclear \& Astrophysics Laboratory, Keble Road, Oxford, OX1 3RH, UK. \\
$^5$ Astrophysics Research Institute, Liverpool John Moores University, 12 Quays House, Egerton Wharf, Birkenhead, CH41 1LD, UK.\\
$^6$ Research School of Astronomy \& Astrophysics, Mount Stromlo Observatory, Institute of Advanced Studies,\\ Australian National University, Private bag, Weston Creek P.O., ACT 2611, Australia.\\
}

\begin{document}
\maketitle
\begin{abstract}

When the 2dF QSO Redshift Survey (2QZ) is complete, a powerful geometric test for the
cosmological constant will be available.
 By comparing the clustering along and across the line of
sight and modelling the effects of peculiar velocities and bulk
motions in redshift space, geometric distortions, which occur if the wrong cosmology is assumed, can be detected. 

In this paper we investigate the effect of geometric and redshift-space distortions in the power spectrum parallel and perpendicular to the observer's line of sight, $P^S(k_{\parallel},\mathbf{k}_{\perp})$.
 Ballinger et al. developed a model to estimate the cosmological constant, $\Lambda$, and the important parameter $\beta \approx \Omega_m^{0.6}/b$ from these distortions. We apply this model to a detailed simulation of the final 25k 2QZ, produced using the Virgo Consortium's huge {\it Hubble Volume} N-body $\Lambda$-CDM light cone simulation. We confirm the conclusions of Ballinger et al.; the shape of the redshift-space and geometric distortions are very similar, and discriminating between the two to produce a purely geometric constraint on $\Lambda$ is difficult. When all the uncertainties in measuring $P^S(k_{\parallel},\mathbf{k}_{\perp})$ for the 2QZ are taken into account we find that only a joint $\Lambda - \beta$ constraint is possible. 

By combining this result with a second constraint based on mass clustering evolution, however, we can make significant progress. We predict that this method should allow us to constrain $\beta$ to approximately $\pm0.1$, and $\Omega_{\rm m}$ to $\pm0.25$ using the final catalogue. We apply the method to the 2QZ catalogue of 10000 QSOs and find that this incomplete catalogue marginally favours a $\Lambda$ cosmology, obtaining best fit values of $\beta = 0.39^{+0.18}_{-0.17}$ and $\Omega_{\rm m} = 1 - \Omega_{\Lambda} = 0.23^{+0.44}_{-0.13}$. 
However, Einstein - de Sitter ($\Omega_{\rm m} = 1.0$, $\Omega_{\Lambda} = 0.0$) models are only rejected at the 1.4$\sigma$ level in the current survey. The rejection of lambda-dominated ($\Omega_{\rm m} = 0.0$, $\Omega_{\Lambda} = 1.0$) models is stronger at $\sim 2\sigma$.

\end{abstract}

\begin{keywords}
surveys - cosmology: observations - large-scale structure of the Universe
\end{keywords}
\section{Introduction}

The large-scale structure of the
Universe represents one of the most powerful discriminants between
cosmological models. We are using the AAT 2dF facility to make a
redshift survey of some 25000 $b_J <20.85^m$ QSOs (Boyle et al. 2000, 
Croom et al. 2001a,b).  QSOs are highly
effective probes of the structure of the Universe over a wide range of
scales and can trace clustering evolution over a look-back time which
is 70-80\%  of its present  age.  

The 2dF QSO Redshift Survey (2QZ)
comprises two $5\times75\deg^2$ declination strips, one at the South
Galactic Pole and one in an equatorial region in the North Galactic
Cap. QSOs are selected by ultra-violet excess (UVX) in the $U$-$B$:$B$-$R$
plane, using APM measurements of UK Schmidt Telescope photographic plates (Smith et al. 2001). The 2dF spectra obtained are identified automatically by \textsc{autoz} (Miller et al., in preparation), which also determines the redshift of the QSOs. These identifications are confirmed by two independent observers.
Approximately 18000 QSO redshifts have been obtained to date,
making this survey already an order of magnitude larger than any
previous QSO survey. 

Observations of high redshift Type Ia Supernovae (SNIa) have recently been used to measure the cosmological constant, $\Lambda$ (e.g. Perlmutter et al. 1999; Reiss et al. 1998). By making the simple assumption that SNIa are standard candles, one can use the apparent luminosity and redshift of such supernovae to constrain the distance-redshift relation, hence determining the value of $\Lambda$. We currently don't have a good understanding of SNIa physical processes, and there are still some uncertainties regarding the standard candle assumption. For example, the above analyses apply a little understood empirical correction to the peak magnitude based on the decline rate of the SN light curve. There have also been suggestions that the SN light curves may be affected by cosmic dust (e.g. Aguirre 1999), or evolve with redshift (e.g. Drell, Loredo \& Wasserman 2000; Reiss et al. 1999a,b; see however Aldering, Knop \& Nugent 2000).

Alcock \& Paczy\'{n}ski (1979) suggested that $\Lambda$ might also be measured from distortions in the shape of large-scale structure. 
Geometrical distortions occur if the wrong cosmology is assumed, due to the different dependence on cosmology of the redshift-distance relation along and across the line of sight. Phillipps (1994) suggested that by making the simple assumption that clustering in real-space is on average spherically symmetric, geometrical distortions due to $\Lambda$ could be detected by comparing the distribution of radial and transverse separations of QSO pairs. This approach has the advantage that it depends only on the assumption of isotropy, without the need for a standard candle. We apply a related test to the 2QZ QSOs. We can measure the clustering of QSOs parallel and perpendicular to the observer's line of sight, through the two-point correlation function, or in this case, the power spectrum, $P^S(k_{\parallel},\mathbf{k}_{\perp})$. Only if we make the correct assumption about geometry would no anisotropic geometric distortion be detected. 

Unfortunately the problem is complicated by the fact that redshift-space distortions are also caused by peculiar velocities. On the large, linear scales probed by the power spectrum coherent peculiar velocities due to infall into overdense regions are the main cause of anisotropy. According to gravitational instability theory these distortions depend only on the density and bias parameters via  $\beta \approx \Omega_{\rm m}^{0.6}/b$. 

In this paper we investigate the redshift-space distortions in the
power spectrum parallel and perpendicular to the observer's line of
sight, $P^S(k_{\parallel},\mathbf{k}_{\perp})$, using mock 2QZ
catalogues generated from the {\it Hubble Volume} simulation (Frenk et al. 2000). The aim is to determine how well we will be able to measure the important cosmological parameters $\Lambda$ and $\beta$ using the 2QZ observations.

We apply the
model from Ballinger et al. (1996) for the redshift-space distortions to a measurement of $P^S(k_{\parallel},\mathbf{k}_{\perp})$ from the simulated QSO catalogue, and derive a joint constraint on $\Lambda$ and $\beta$. 
The angular dependence of geometrical distortions differs only slightly from that of the redshift-space distortions, and so the constraint is fairly degenerate. To break the degeneracy, different constraints on $\Lambda$ and $\beta$ need to be considered. Here we examine one possibility based on the evolution of mass clustering from the average redshift of the 2QZ to the present day.  Finally we apply the method to the first public release of the 2QZ; the {\it 10k Catalogue} (Croom et al. 2001b).

\section{Power Spectrum Analysis}

\subsection{Mock 2dF QSO Catalogues}

In order to test how effectively the 2QZ will measure
large-scale structure, we are using the Virgo Consortium's huge {\it Hubble Volume}
N-body $\Lambda$CDM simulation (Frenk et al. 2000) to simulate three light cone strips of the 2QZ. The parameters of the simulation are $\Omega_{\rm b}$=0.04, $\Omega_{\rm CDM}$=0.26, $\Omega_{\Lambda}$=0.7, $H_{\circ}$=70 km s$^{-1}$Mpc$^{-1}$ and the normalisation, $\sigma_8$, is 0.9, consistent with the abundance of hot X-ray clusters (White, Efstathiou \& Frenk 1993) and with the level of anisotropies in the cosmic microwave background radiation (CMB) found by COBE (Smoot et al. 1992). One billion mass particles are contained within a cube that is 3,000$h^{-1}$Mpc on a side. One of the vertices was chosen to be the observer and the long axis of the light-cone was oriented along the maximal diagonal. The light-cone therefore extends to a depth of $\sim$5,000$h^{-1}$Mpc, which corresponds to $z \sim $4 in the $\Lambda$CDM cosmology. The solid angle of the light-cone is 75$\times$15 degrees which is split into three 75$\times$5 degree slices. The 2QZ consists of 2 such slices, and so to generate the mock catalogue we use the two disconnected slices. We also consider the third slice for the error analysis.

To create realistic mock catalogues, we must bias the mass particles to give a similar clustering pattern, and the same angular and radial selection function and sampling expected in the final 2dF Survey. 
The biasing prescription adopted follows method 2 as described by Cole et al. (1998), with the parameters chosen to produce a constant clustering amplitude with redshift - as seen in previous surveys (Croom \& Shanks 1996). It is an Eulerian, or final density scheme due to the nature of the N-body simulation output, and is based only on the local density field, implying asymptotic scale-independence of the bias on the large scales we are probing. The biased particles were then sparsely sampled to match the radial distribution of the QSOs in the preliminary 2dF catalogue, and the expected number density of the final catalogue over the redshift range $0.3 < z < 2.2$. For further details see Hoyle (2000).

The {\it Hubble Volume} simulation was designed to probe large-scale structure, at the expense of spatial resolution. As a result the simulation does not probe any structure on scales $\la 3\;h^{-1}\;$Mpc. In this analysis we are binning the data onto a grid with bin size $\sim 12\;h^{-1}\;$Mpc, and only considering power on scales $\ga 32\;h^{-1}\;$Mpc, so we expect this to have little effect on our final results.

\subsection{Power spectrum estimation}

\begin{figure}
\centerline{\hbox{\psfig{figure=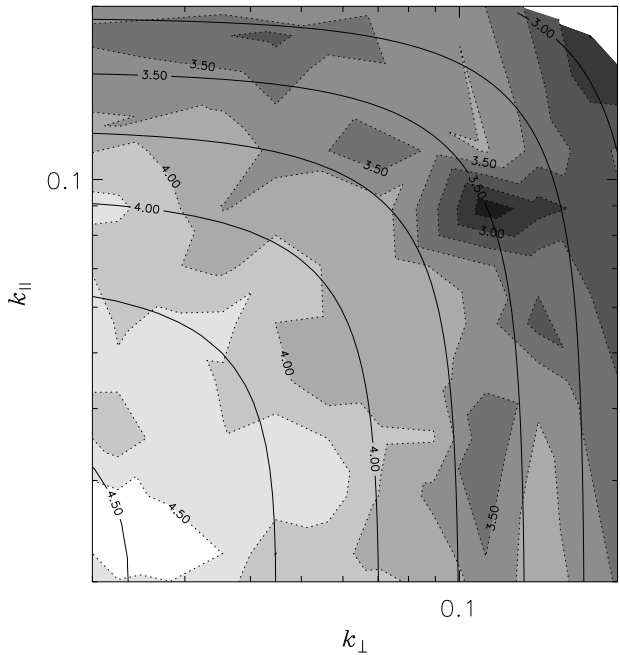,width=9.5cm}}}
\caption{$P^S(k_{\parallel},\mathbf{k}_{\perp})$ estimated from the 25k mock QSO survey. Filled contours of constant ${\mathrm log} (P(k)/h^{-3}$Mpc$^3)$ are shown as a function of $k_{\parallel}/h$\,Mpc$^{-1}$ and ${\mathbf k}_{\perp}/h$\,Mpc$^{-1}$. Overlaid is the best-fit model (see section~\ref{massclu}) for $P^S(k_{\parallel},\mathbf{k}_{\perp})$, with a flat $\Omega_{\rm m} = 0.33$ universe, with $\beta = 0.33$ and $\sigma_p = 400\;$km$\;$s$^{-1}$.
}
\label{fig2}
\end{figure}

The power spectrum estimation is carried out as described in Outram, Hoyle \& Shanks (2001), using the method outlined in Hoyle et al. (1999)
 and Tadros \& Efstathiou (1996). In order to apply the distant observer
 approximation (see section~\ref{datageom}) the mock QSO survey is
 divided into four regions. Each region is embedded
 into a larger cubical volume, which is rotated  such that the central
 line of sight lies along the same axis of the cube in each case. The
 density field is binned onto a 256$^3$ mesh, using nearest grid-point
 assignment.  The power spectrum of each region is estimated using a
 Fast Fourier Transform (FFT), and the average of the resulting power
 spectra is taken. The result, binned logarithmically into
 $k_{\parallel}$ and $\mathbf{k}_{\perp}$, is plotted in figure~\ref{fig2}.

\subsection{Survey geometry}\label{datageom}
The 2QZ covers two declination strips of approximately $5^{\circ} \times 75^{\circ}$, and we have three such mock declination strips available.
When calculating $P^S(k_{\parallel},\mathbf{k}_{\perp})$ information about the line of sight must be retained. To achieve this the data are divided into subsamples that subtend a small solid angle on the sky, and the distant-observer approximation is then applied. The data from each declination strip are split into 4 regions each approximately $5^{\circ} \times 20^{\circ}$ during this analysis, and the average of the resulting power spectra is taken. 

This approach introduces a small systematic error; the measured $P^S(k_{\parallel},\mathbf{k}_{\perp})$ is in fact a convolution between the true power spectrum and a function related to the sample window introduced due to the assumption of parallel lines of sight. The resulting power spectrum is more isotropic, and this leads to a small, yet systematic under-estimate of $\beta$.  For a $5^{\circ} \times 20^{\circ}$ window this error is $< 1\%$ and considerably smaller than the statistical errors obtained. Although it is possible to correct for this effect (see Cole et al. 1994), for the purposes of this analysis we have chosen to ignore it.

\begin{figure}
\centerline{\hbox{\psfig{figure=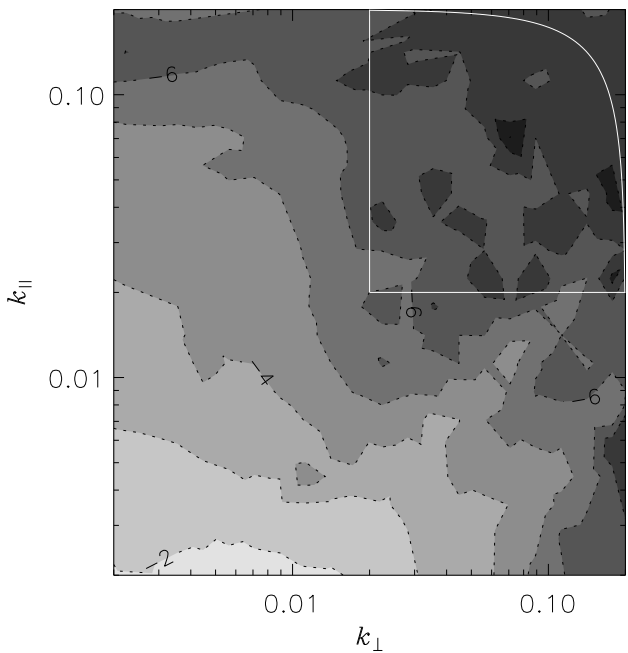,width=9.5cm}}}
\caption{The power spectrum of the 25k mock QSO survey window function.
Filled contours of constant ${\mathrm log} (|\hat{W}|^2/h^{-3}$Mpc$^3)$ are shown as a function of $k_{\parallel}/h$\,Mpc$^{-1}$ and ${\mathbf k}_{\perp}/h$\,Mpc$^{-1}$. To minimise the effects of the window function, we consider only the region bound by the white line above, with $k_{\parallel}$ \& $|\mathbf{k}_{\perp}|$ $ \ge0.02\;h\;$Mpc$^{-1}$.}
\label{fig1}
\end{figure}
 
\begin{figure}
\centerline{\hbox{\psfig{figure=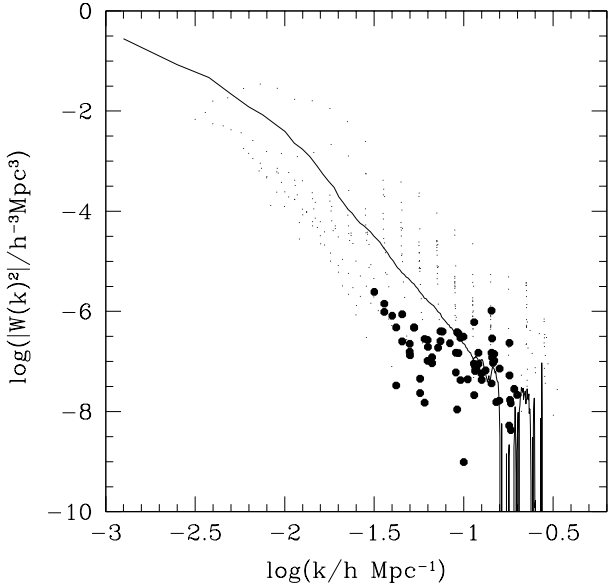,width=9.5cm}}}
\caption{The power spectrum of the 25k mock QSO survey window function (points), shown as a function of $k$. The bold points are those which were used in the analysis (see Fig.~\ref{fig1}), with wavenumbers with $k_{\parallel}>0.02, |{\mathbf k}_{\perp}|>0.02$, and $k<0.2$. 
The window function  is a steep power law, varying as $\sim k^{-3}$.
  The overlaid line is the window function from the usual power spectrum analysis, binning in shells of $k$, of the 25k mock catalogue.
}
\label{figwin1d}
\end{figure}

The measured power spectrum is also convolved with the power spectrum of the window function. As the QSO co-moving number density is almost constant out to very large scales (Hoyle et al. 2001a), we are effectively considering a volume limited sample. Hence each QSO carries equal weight, and the survey window function, $W({\mathbf x})$, simply takes a value of unity in the volume of the universe included in the survey, and zero elsewhere. This is approximated using a catalogue containing a large number of unclustered points with the same radial and angular distribution as the survey, and its power spectrum is calculated in a similar manner to that of the data. Figure~\ref{fig1} shows the power spectrum of the window function, $|\hat{W}|^2$, as a function of $k_{\parallel}$ and $\mathbf{k}_{\perp}$, and figure~\ref{figwin1d} shows the window function plotted as a function of $k$, with a comparison against that of a normal $P(k)$ analysis, binning in shells of $k$.

The window function is sharply peaked in ${\mathbf k}$-space, varying as a steep power law,  $\sim k^{-3}$ or steeper; much steeper than the expected QSO power spectrum. Therefore the convolution should have little effect at wavenumbers where $|\hat{W}|^2$ takes a value much less than about a tenth of its maximum value. There is some evidence in figure~\ref{fig1} that the window function is slightly shallower along, and in particular across the line of sight. The directional dependence of the window function also explains the scatter seen in figure~\ref{figwin1d}. This extension at low values of $k_{\parallel}$ is due to the relatively narrow width of the declination strips; $5^{\circ}$ at $z=2.4$ corresponds to $r \sim 300 h^{-1}\;$Mpc, or $k\sim 0.02$. To avoid any problems caused by such effects we consider only those regions with $k_{\parallel}$ \& $|\mathbf{k}_{\perp}|$ $ \ge0.02\;h\;$Mpc$^{-1}$. 

\subsection{Error analysis}

\begin{figure}
\centerline{\hbox{\psfig{figure=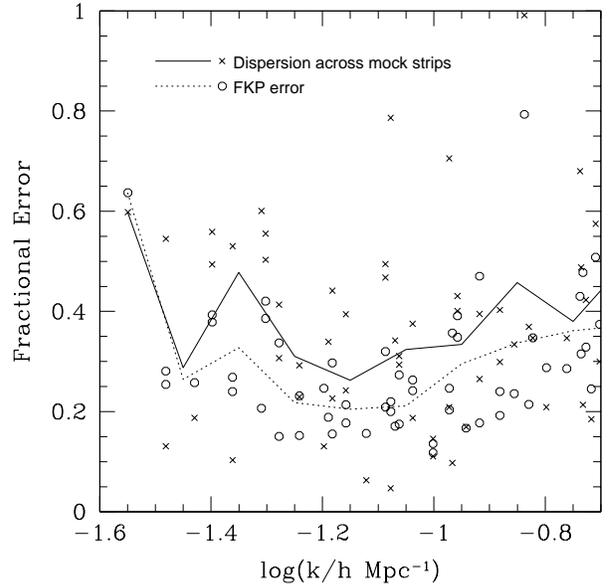,width=9.5cm}}}
\caption{A comparison of the fractional errors using FKP estimation (circles) and the dispersion across 3 mock strips (crosses). The dotted and solid lines show the average error, binned in shells of $k$, calculated using FKP estimation and the  dispersion across 3 mock strips respectively.
}
\label{figerr}
\end{figure}

\begin{figure}
\centerline{\hbox{\psfig{figure=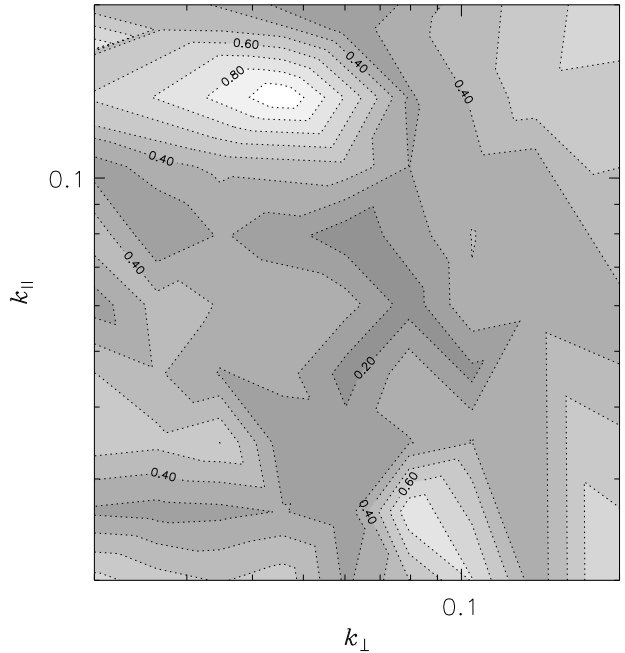,width=9.5cm}}}
\caption{The adopted fractional errors in $P^S(k_{\parallel},\mathbf{k}_{\perp})$ (the larger of the FKP error and the dispersion across the three catalogues in each bin) are shown as a function of $k_{\parallel}/h$\,Mpc$^{-1}$ and $\mathbf{k}_{\perp}/h$\,Mpc$^{-1}$. The fractional errors vary from $<0.2$ in the darkest region to $\sim0.8$ in the lightest region.
}
\label{figerr2}
\end{figure}

The errors in the power spectrum are estimated in two ways. Firstly we have three separate strips on which the analysis is carried out, and we can estimate the error from the dispersion in the three measurements. 
Alternatively we can consider the error estimate of Feldman, Kaiser \& Peacock (1994; FKP). A comparison of the two error estimates is shown in figure~\ref{figerr}. The two estimates agree fairly well, but with large scatter. Although some of the scatter in the error estimates is intrinsic, the dependence of the error on the angle of the wavevector to the line of sight, as seen in figure~\ref{figerr2}, also contributes to the scatter. The FKP error is, on average, slightly lower than the error estimated from the dispersion across the mock strips.  Neither approach is ideal; the error estimated from the dispersion has a high uncertainty due to the small number of mock catalogues, whereas the FKP error estimate assumes a thin shell geometry, and can under-estimate the error on some scales, due, for example, to the onset of non-linearity. To minimise such problems, we took the error in each bin to be the larger of the two estimates. The adopted fractional errors are shown in figure~\ref{figerr2}. 

\section{Modelling Redshift Distortions}\label{secmod}
The power spectrum model incorporating redshift distortions that we fit to the data is presented in Ballinger, Peacock and Heavens (1996). For clarity and convenience we review the main details below.

\subsection{Geometrical distortions}

The power spectrum analysis was carried out assuming an $\Omega_{\rm m}$=0.3, $\Omega_{\Lambda}$=0.7 universe, matching that of the {\it Hubble Volume} simulation. If the true geometry differs from this then our distance calculations will be wrong be a factor $f_{\perp}$ perpendicular to the line of sight, and $f_{\parallel}$ along the line of sight (as defined in Ballinger et al. 1996). Thus we can define a geometric flattening factor:

\begin{equation}\label{flateqn}
F(z)=\frac{f_{\parallel}}{f_{\perp}}
\end{equation}
(Ballinger et al. 1996). For an Einstein - de Sitter ($\Omega_{\rm m} = 1.0$, $\Omega_{\Lambda} = 0.0$), and lambda-dominated ($\Omega_{\rm m} = 0.0$, $\Omega_{\Lambda} = 1.0$) universe, $F \approx 0.81$ and $1.53$ respectively at the mean depth of the survey ($z \approx 1.4$). The effect this has on the power spectrum is given by
\begin{equation}\label{geomeqn}
P_{\mathrm anisotropic}(k_{\parallel},{\mathbf k}_{\perp})= 
\frac{P_{\mathrm true}(k)}{f_{\perp}^{3+n}F}  
\left[1+\mu^2\left(\frac{1}{F^2}-1\right)\right]^{\frac{n}{2}}
\end{equation}
(Ballinger et al. 1996) where $\mu=k_{\parallel}/k$, and $n$ is the spectral index of the power spectrum. 
\subsection{Redshift distortions}\label{reddist}
On large, linear scales the main cause of anisotropy are coherent peculiar velocities due to the infall of galaxies into overdense regions. This anisotropy takes a very simple form in redshift-space, depending only on the density and bias parameters via the combination $\beta \approx \Omega_{\rm m}^{0.6}/b$:
\begin{equation}\label{betaeqn}
P^S(k_{\parallel},{\mathbf k}_{\perp}) =
P^R(k)\left[1+\beta\mu^2\right]^2
\end{equation}
(Kaiser 1987) where $P^S$, and $P^R$ refer to the redshift-space and real-space power spectra respectively. The Kaiser formula also assumes the distant-observer approximation, and hence is valid in this case due to the method of estimation of $P^S(k_{\parallel},{\mathbf k}_{\perp})$ (\S~\ref{datageom}).

We expect the parameters $b$, $\Omega_{\rm m}$, and hence $\beta$ to vary as a function of redshift. To keep the number of free parameters to a minimum, however, we have chosen to fit a single average value of $\beta_{\rm QSO}(z\sim1.4)$ over the redshift range, $0.3<z<2.2$, probed by the QSOs. Large variations of $\beta$ over this range could introduce a systematic error into our results, however, the value of $\beta(z\sim1.4)$ determined from the first 10000 2dF QSOs suggests that any variation with redshift is relatively small (see section~\ref{sec5}). We hope to investigate the evolution of $\beta$ in more detail using the final 2QZ sample of 25000 QSOs.

Anisotropy in the power-spectrum is dominated by the galaxy velocity dispersions in virialized clusters on small, non-linear scales. This is modelled by introducing a damping term. The line of sight pairwise velocity ($\sigma_p$)\footnote{In power spectra $\sigma_p$ is implicitly divided by $H_0$ and quoted in units $h^{-1}\;$Mpc. ($H_0=100h\;$km$\;$s$^{-1}\;$Mpc$^{-1}$) } distribution is modelled using an exponential, which produces a better fit to both observations and N-body simulations than either a Gaussian or power law distribution (Davis \& Peebles 1983; Fisher et al. 1994). Observational redshift determination errors may add a second, Gaussian component to the observed velocity dispersion of the 2QZ QSOs (see section~\ref{rederr}). An exponential is given by a Lorentzian factor in redshift-space:
\begin{equation}
D\left[k\mu\sigma_p\right]=\frac{1}{1+\frac{1}{2}\left(k\mu\sigma_p\right)^2}
\end{equation}
This model is only a good approximation in mildly non-linear situations. Cole et al. (1995) used numerical simulations to show that it is effective for wavelengths $\lambda \ga 15h^{-1}\;$Mpc, or $k\la0.4$; considerably smaller than the scales we are considering. At these wavenumbers the difference between models assuming a Gaussian or exponential velocity dispersion is negligible (Ballinger et al. 1996). 

Combining these effects leads to the final model:
\begin{eqnarray}\label{model}
P^S(k_{\parallel},{\mathbf k}_{\perp}) &= &
\frac{P^R(k)}{f_{\perp}^{3+n}F}  
\left[1+\mu^2\left(\frac{1}{F^2}-1\right)\right]^{\frac{n-4}{2}}\nonumber\\
&\times& \left[1+\mu^2\left(\frac{\beta+1}{F^2}-1\right)\right]^2
D\left[k\mu\sigma_p^{\prime}\right]
\end{eqnarray}
(Ballinger et al. 1996) where $\sigma_p^{\prime}=\sigma_p/f_{\parallel}$.

\begin{figure}
\centerline{\hbox{\psfig{figure=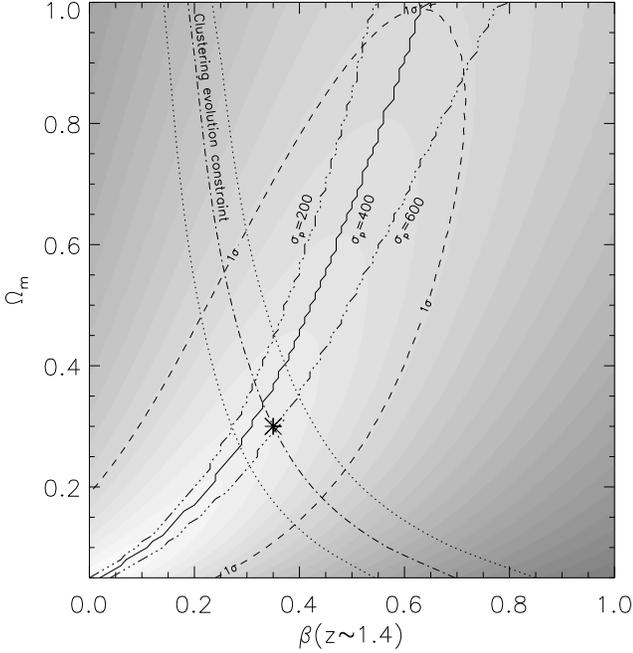,width=9.5cm}}}
\caption{Filled contours of increasing $\chi^2$ in the $\Omega_{\rm m}$ -- $\beta$ plane for the 25k mock QSO survey, assuming $\sigma_p=400\;$km$\;$s$^{-1}$. The solid line represents the best fit value of $\beta$ for each $\Omega_{\rm m}$ and the 1-$\sigma$ error is given by the dashed line. The dot-dot-dot-dash lines above and below the solid line show the  best fit value of $\beta$ for each $\Omega_{\rm m}$ with $\sigma_p=200\;$km$\;$s$^{-1}$ and $\sigma_p=600\;$km$\;$s$^{-1}$ respectively.
Overlaid is the best-fit (dot-dash) and 1-$\sigma$ (dot) values of  $\beta$ determined using the method described in section~\ref{massclu}
The joint best fit values obtained are  $\beta = 0.33$ and $\Omega_{\rm m} = 0.33$. The star marks the input {\it Hubble Volume} model values of $\beta_q = 0.35$ and $\Omega_{\rm m} = 0.3$.
}
\label{fig3}
\end{figure}

There are several free parameters in the model; $F, \beta, \sigma_p$, and $P^R(k)$, the underlying real-space power spectrum. To reduce the uncertainties on the parameters of interest, $\beta, \& F$ we make simple assumption about the other parameters. $\sigma_p$ has relatively little effect on the power spectrum at large scales and so is not well constrained in this analysis. We have chosen to fix it at $400\;$km$\;$s$^{-1}$; the measured value of the {\it Hubble Volume} simulation at $z\sim1.4$ (Hoyle 2000). This is similar to recent measurements from local galaxy surveys (Outram et al. 2001, Peacock et al. 2001). Whilst it may be expected that $\sigma_p$ falls with increasing redshift, the finite resolution of the {\it Hubble Volume} simulation, $\sim 3\;h^{-1}\;$Mpc, introduces an additional velocity dispersion of similar order that keeps the observed value artificially high. A similar effect due to redshift determination errors affects the true QSO sample and is discussed later. To test any systematics introduced by this assumption we repeat the analysis with $\sigma_p=200\;$km$\;$s$^{-1}$ and $\sigma_p=600\;$km$\;$s$^{-1}$. The effects can be seen in figure~\ref{fig3}. There is a small systematic shift in the best-fit $\beta - \Omega_{\rm m}$ relation that becomes insignificant at low values of $\Omega_{\rm m}$. A similar analysis of redshift-space distortions considering $\xi(\sigma,\pi)$ (Hoyle et al. 2001b) is sensitive to smaller-scale clustering and so should constrain $\sigma_p$ in the 2QZ more effectively.

We reconstruct a real-space power spectrum self-consistently from the measured mock QSO redshift-space power spectrum (Hoyle et al. 2001a), by inverting the above redshift distortion equations in each cosmology for an average value of $\mu$. Although the input real-space power spectrum is available for the simulation, this is not the case for the 2QZ, and so we choose not to use it. Finally, we assume a flat cosmology, consistent with the recent CMB results from experiments such as Boomerang (de Bernardis et al. 2000) and Maxima (Balbi et al. 2000). We choose to fit the variable $\Omega_{\rm m}$, and fix $\Omega_{\Lambda}=1-\Omega_{\rm m}$.

\begin{figure}
\centerline{\hbox{\psfig{figure=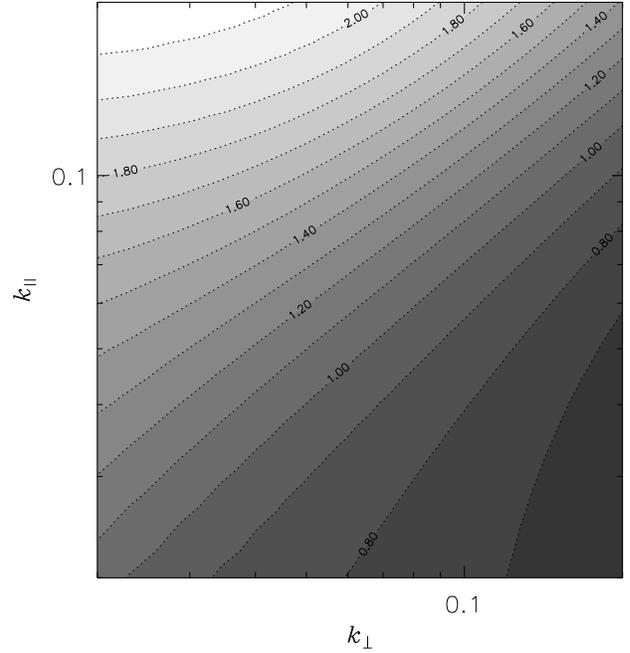,width=9.5cm}}}
\caption{The ratio of two power spectrum models, $\beta = 0.39$, $\Omega_{\rm m} = 0.23$, and $ \beta = 0.19$, $\Omega_{\rm m} = 1.0$ (shown in Fig.~\ref{fig10000data}). This figure illustrates how the fractional anisotropy in $P^S(k_{\parallel},\mathbf{k}_{\perp})$ varies with cosmology. To constrain $\beta$ and $\Lambda$, we need to detect variations of order two in the power spectrum anisotropy caused by geometrical and redshift distortions. 
}

\label{modrat}
\end{figure}

Figure~\ref{modrat} illustrates how the fractional anisotropy in $P^S(k_{\parallel},\mathbf{k}_{\perp})$ varies with cosmology. The ratio of two power spectrum models, with $\beta = 0.39$, $\Omega_{\rm m} = 0.23$, and $ \beta = 0.19$, $\Omega_{\rm m} = 1.0$ (shown in Fig.~\ref{fig10000data}) are plotted. To discriminate between these two models, and hence constrain $\beta$ and $\Lambda$, we need to detect variations of order two in the power spectrum due to geometrical and redshift distortions. 

\subsection{$\chi^2$ fitting}

We fit the model to the data using a $\chi^2$ technique:
\begin{equation}
\chi^2=\sum\frac{( P_{\mathrm observed}- P_{\mathrm model})^2}{\sigma_{ P(k)}^2}
\end{equation}
The errors, $\sigma_{ P(k)}$, are the larger of the FKP error and the dispersion across the three catalogues in each bin. Only those wavenumbers with $k_{\parallel}>0.02, |{\mathbf k}_{\perp}|>0.02$, and $k<0.2$ are used in the fit. The former constraints are applied to remove the effects of the window function, and the latter because the FFT is unreliable at smaller scales. The Nyquist frequency of the transform is $k=0.26$. This constraint also prevents excessive non-linearity, where the model breaks down. 

The $\beta$, and $\Omega_{\rm m}$ parameters in the model are adjusted until a minimum value of $\chi^2$ is obtained. Figure~\ref{fig3} shows $\chi^2$ contours in the $\Omega_{\rm m}$ -- $\beta$ plane. Nominally the best fit value is $\beta=0.02$, and $\Omega_{\rm m}=0.05$, with $\chi^2=98.46$ over 69 degrees of freedom. This suggests that the errors have been slightly under-estimated. However, as figure~\ref{fig3} shows, the fit is highly degenerate, and any values along the solid line are acceptable. The joint best fit value (see below) of $\beta = 0.33$ and $\Omega_{\rm m} = 0.33$ has $\chi^2=100.32$, only marginally higher.

\section{Mass Clustering Evolution}\label{massclu}

Applying the
model from Ballinger et al. (1996) for the redshift-space distortions to a measurement of $P^S(k_{\parallel},\mathbf{k}_{\perp})$ from the simulated QSO catalogue has given us a joint constraint on $\Omega_{\rm m}$ and $\beta$.
The angular dependence of geometrical distortions differs only slightly from that of the redshift-space distortions, and so the constraint is fairly degenerate. To break this degeneracy, a different constraint on $\Omega_{\rm m}$ and $\beta$ needs to be considered. Here we examine one possibility based on the evolution of mass clustering from the average redshift of the 2QZ to the present day.

 One of the early results from the 2dF Galaxy Redshift Survey is a measurment of $\beta_{\rm g}(z\sim0) = 0.43\pm0.07$ (Peacock et al. 2001), using redshift-space distortions in the two-point galaxy correlation function. For each cosmology (again we only consider flat cosmologies) the value of the galaxy-mass bias can be found from $\beta_{\rm g}$, which in turn gives the value of the mass correlation function if the galaxy correlation function at $z=0$ is known, as shown below. The evolution in the mass correlation function can be calculated for this cosmology, and hence by comparing the $z\sim1.4$ to mass correlation function to that of the QSOs, an estimate of the QSO bias factor can be obtained. Finally, this can be used to derive an estimate of $\beta_q$ as a function of cosmology. One possible caveat with this technique is that we are comparing values of $\beta_g$ and $\beta_q$ that were measured using different estimators and on different scales. To minimise any possible systematic effects this analysis could be carried out self-consistently using the same method to determine $\beta_g$ (Outram, Hoyle \& Shanks 2000). As the 2dF Galaxy Redshift Survey data is not yet available, we have chosen for simplicity to use a correlation function analysis.

Rather than determining the value of the two-point correlation function at one particular point, we use the less noisy volume averaged two-point correlation function out to $20 h^{-1}$Mpc, $\bar{\xi} (20)$. To estimate the redshift space, volume averaged two-point correlation function of galaxies, $\bar{\xi}^s_g$, we assume that the galaxy correlation function can be approximated by a power law of the form $\xi^s_g=(s/6 h^{-1}$Mpc$)^{-1.7}$. The power law approximation is in very good agreement with early results from the 2dF Galaxy Redshift Survey two-point correlation function over the range of scales 2$<s<$20$h^{-1}$Mpc (Peder Norberg, private communication). Once the 2dF Galaxy Redshift Survey is completed, there will be no need to make this approximation.
$\bar{\xi}_g$ is then found via
\begin{equation}
\bar{\xi}^s_g (20) = \frac{ \int_0 ^{20} \xi^s_g 4 \pi s^2  ds}{\int_0 ^{20} 4 \pi s^2 ds} = 0.3.
\label{eq:int}
\end{equation}
By integrating out to 20$h^{-1}$Mpc, the non-linear effects on the volume averaged correlation function should be small, and are not considered in this analysis. Redshift measurement errors should only be a factor on small, $\la 5 h^{-1}$Mpc, scales and so are also not considered.

For each cosmology, the bias between the galaxies and the mass at $z$=0, $b_{g \rho}(0)$, can be found from
\begin{equation}
b_{g\rho}(0) = \frac{\Omega_{\rm m}^{0.6}(0)} {\beta_{\rm g}(0)} .
\end{equation}
The real space galaxy correlation function can be determined from the redshift space galaxy correlation function
\begin{equation}
\bar{\xi}^r_g = \frac{\bar{\xi}^s_g}{[1 + \frac {2}{3} \beta_{\rm g}(0) + \frac{1}{5} \beta^2_{\rm g}(0)]},
\end{equation}
where the superscripts $r$ and $s$ indicate real and redshift space respectively.
The real space mass correlation function at $z$=0 can now be found as
\begin{equation}
\bar{\xi}^r_{\rho}(0) = \frac{\bar{\xi}^r_g(0)}{b_{g\rho}^2(0)}.
\end{equation}
The real space mass correlation function evolves according to linear theory such that
\begin{equation}
\bar{\xi}^r_{\rho}(z) = \frac{\bar{\xi}^r_{\rho}(0)} {G(z)^2},
\label{eq:evo}
\end{equation}
where $G(z)$ is the growth factor, which depends on cosmology, found from the formula of Carroll, Press \& Turner (1992). When $\Omega_{\rm m}$=1, $G(z) = (1 + z)$.

We now relate the correlation function of the mass at $z$ measured in real space to the amplitude of the QSO clustering at $z$, measured in redshift space as we wish to know $\beta_q(z)$, where the subscript $q$ stands for QSO, as a function of $\Omega_{\rm m}(0)$. 
$\bar{\xi}^s_q (20)$, measured over the redshift range $0.3<z<2.2$, is determined separately in each cosmological model from the latest 2QZ results (see Croom et al. (2001a) for details). In this analysis we have not taken into account the effect of redshift determination errors on the measurement of the QSO correlation function. This may lead to a slight systematic overestimate of the QSO bias.

First we calculate $\Omega_{\rm m}(z)$ using
\begin{equation}
\Omega_{\rm m}(z) = \frac {\Omega_{\rm m}(0) (1 + z)^3} {\Omega_{\rm m}(0) (1 + z)^3 + \Omega_{\Lambda} }
\end{equation}
(valid for flat cosmologies only). $\beta_q(z)$ is then given by
\begin{equation}
\beta_q(z) = \frac { \Omega_{\rm m}(z)^{0.6}} {b_{q\rho}(z)},
\label{eq:betaq}
\end{equation}
where $b_{q\rho}(z)$ is defined by
\begin{equation}
b^2_{q\rho}(z) = \frac{\bar{\xi}^r_q(z)}{\bar{\xi}^r_{\rho}(z)}.
\label{eq:betaq2}
\end{equation}
$\bar{\xi}^r_q(z)$ can be found via 
\begin{equation}
\bar{\xi}^r_q(z) = \frac{\bar{\xi}^s_q(z)}{(1 + \frac{2}{3} \beta_q(z) + \frac{1}{5} \beta^2_q(z)) }.
\label{eq:xiqrxiqs}
\end{equation}
As $\bar{\xi}^s_q$ is remeasured in each cosmological model, this is valid without consideration of geometric distortions. Combining equations \ref{eq:betaq},\ref{eq:betaq2}, and \ref{eq:xiqrxiqs} gives
\begin{equation}
\beta^2_q(z) = \Omega_{\rm m}(z)^{1.2} \frac { \bar{\xi}^r_{\rho}(z) }{ \bar{\xi}^s_q(z)} [1 + \frac{2}{3} \beta_q(z) + \frac{1}{5} \beta^2_q(z)],
\end{equation}
a quadratic in $\beta^2_q(z)$ which can easily be solved, allowing $\beta_q(z)$ for any $\Omega_{\rm m}(0)$ to be found. We substitute $z=1.4$ to find the value of $\beta_q(z)$ at the average redshift of the survey. 

\begin{figure}
\centerline{\hbox{\psfig{figure=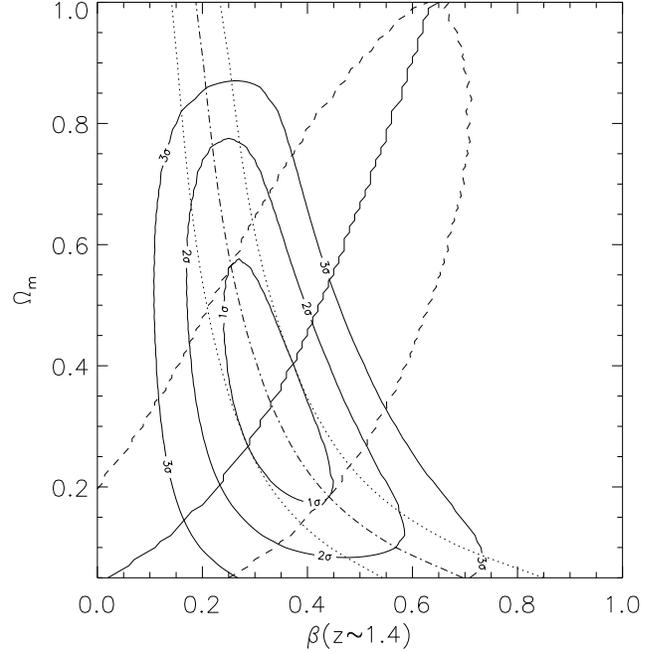,width=9.5cm}}}
\caption{Significance contours given by joint consideration of the two constraints from the 25k {\it Hubble Volume} mock catalogues.
The joint best fit values obtained are  $\beta = 0.33$ and $\Omega_{\rm m} = 0.33$.}
\label{figjoint}
\end{figure}

The errors on $\beta_q(z)$ are found in the standard way, i.e. by differentiating $\beta_q(z)$ with respect to all the variables that contribute to the error and summing the components of the error in quadrature. In this case there are errors on $\beta_g(0)$, $\bar{\xi}^s_q(z)$ and $\bar{\xi}^s_g(0)$. The error on $\beta_g(0)$ is $\pm$ 0.07, the error on $\bar{\xi}_q(z)$ is the Poisson error found from the total $DD$ counts. The error on 
$\bar{\xi}_g$ is the least well known as the power law approximation is used. We assume an error of around 20\%, which is larger than the error expected from the completed 2dF Galaxy Redshift Survey. 

To produce a fully self-consistent constraint on the mock catalogue data, we need to consider measurements of $\beta_g(0)$ \& $\bar{\xi}_g^s$ for the galaxies and $\bar{\xi}^s_q$ for the QSOs from the same simulation, rather than real data. To do this properly would require producing a mock galaxy catalogue. Alternatively we can appeal to the fact that the value of $\bar{\xi}^r_{\rho}(z=1.4)$ in an $\Omega_{\rm m}$=0.3, $\Omega_{\Lambda}$=0.7 universe, obtained above, is in very good agreement with that of the {\it Hubble Volume}. Also, the {\it Hubble Volume} QSO mock catalogue was designed to have a correlation function that closely matches the observed 2dF QSO clustering properties (Hoyle 2000). Hence the constraint we derive below is still valid.

The value of $\beta_q(z)$ and the one-sigma errors are plotted on figure~\ref{fig3}. Although totally degenerate with the value of $\Omega_{\rm m}$, this provides a different constraint to that from the redshift-space distortions. This was derived using significantly smaller scales than the $P^S(k_{\parallel},\mathbf{k}_{\perp})$ analysis, and so we can treat the results as independent and hence combine the likelihoods, yielding a much stronger fit.
The joint best fit values obtained are  $\beta_q = 0.33\pm0.10$ and $\Omega_{\rm m} = 0.33\pm0.22$. The one, two and three-sigma significance contours are plotted in figure~\ref{figjoint}. These results are consistent with the input values of $\beta_q = 0.35$ and $\Omega_{\rm m} = 0.3$, and an Einstein-de Sitter  model is ruled out in the {\it Hubble Volume} at over three-sigma.

\section{The 2QZ 10k Catalogue}\label{10ksec}

\begin{figure}
\centerline{\hbox{\psfig{figure=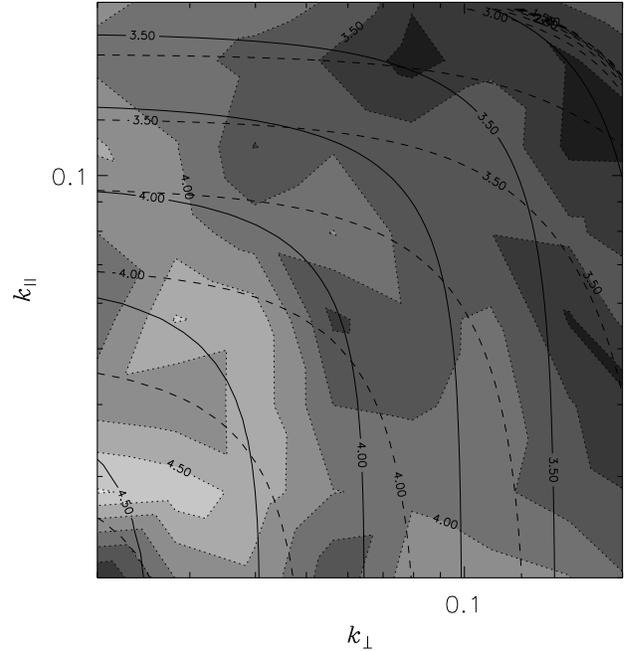,width=9.5cm}}}
\caption{$P^S(k_{\parallel},\mathbf{k}_{\perp})$ estimated from the 2QZ 10k Catalogue containing 10681 QSOs. Filled contours of constant ${\mathrm log} (P(k)/h^{-3}$Mpc$^3)$ are shown as a function of $k_{\parallel}/h$\,Mpc$^{-1}$ and ${\mathbf k}_{\perp}/h$\,Mpc$^{-1}$. Overlaid are the best fit model (solid contours) with $\beta = 0.39$ and $\Omega_{\rm m} = 0.23$, and a model with $\Omega_{\rm m} = 1.0$ and $\beta = 0.19$ (dashed contours).
}
\label{fig10000data}
\end{figure}

We now apply this analysis to the 2QZ 10k Catalogue (Croom et al. 2001b). This catalogue contains the most spectroscopically complete 2dF fields observed prior to November 2000. The colour selection technique used to create the input catalogue of QSO candidates for identification using the 2dF instrument gives high completeness ($>$90\%) out to $z\sim2.2$ (Boyle et al. 2000). Adopting this as our high redshift limit to ensure high completeness, we use a sample of 8935 QSOs with redshift $0.3<z<2.2$ in the following power spectral analysis.

The measured QSO power spectrum is a convolution of the intrinsic power spectrum with that of the window function (see section~\ref{datageom}). As the survey is still incomplete, it currently has a patchy angular selection function and a completeness map has to be constructed. This is used to create the random catalogue that approximates the complicated window function in the analysis. Extinction due to galactic dust was taken into account using the maps of Schlegel, Finkbeiner \& Davis (1998). The construction of the window function is described in more detail in Hoyle et al. (2001a). 

There is a small anisotropy introduced by the minimum separation of fibres on the 2dF instrument. As the fibres cannot be placed closer together than $30''$, close pairs of QSOs cannot be observed simultaneously, and this introduces an angular selection function against such pairs in the 10k catalogue. This is being addressed through separate observations of the QSOs missed for the final catalogue. The scales we are probing are considerably larger than those affected by this selection effect; at $z\sim1.4$, $k\la0.2$ corresponds to $\theta\ga40'$.
 Hence we expect that this selection function should have little effect on our results.

\begin{figure}
\centerline{\hbox{\psfig{figure=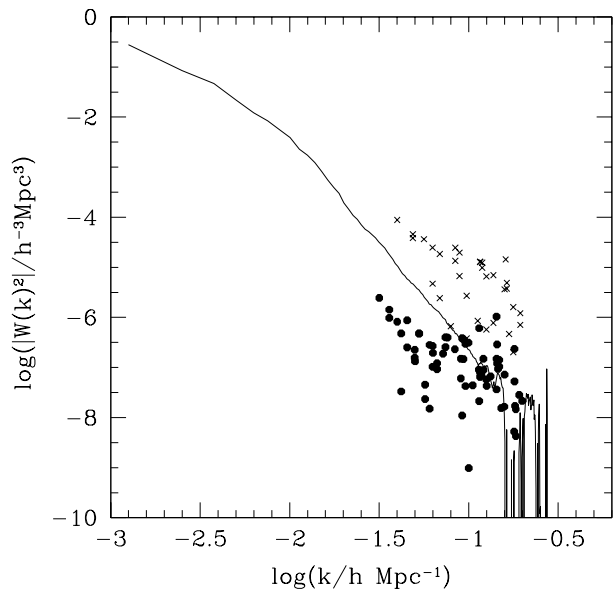,width=9.5cm}}}
\caption{The power spectrum of the window function of the 2QZ 10k catalogue, shown as a function of $k$. The crosses  show the points which were used in the 10k analysis, with wavenumbers with $k_{\parallel}>0.02, |{\mathbf k}_{\perp}|>0.02$, and $k<0.2$. 
The window function  is a steep power law, varying as $\sim k^{-3}$.
The points show the power spectrum of the window function from the analysis of the 2QZ 25k  mock catalogue. The 10k catalogue points are slightly higher than those of the full mock catalogue due to the current incomplete coverage of the survey.  The overlaid line is the window function from the usual power spectrum analysis, binning in shells of $k$, of the 25k mock catalogue.
}
\label{figwin1d10}
\end{figure}

Due to the lower number of QSOs, and a more complicated window function, the power spectrum measurements were very noisy, and only coarse binning in $k_{\parallel}$ and $\mathbf{k}_{\perp}$ was possible.
Figure~\ref{fig10000data} shows $P^S(k_{\parallel},\mathbf{k}_{\perp})$ estimated from the 2QZ 10k catalogue.
The power spectrum of the window function of the 10k catalogue are shown in
figure~\ref{figwin1d10}. Due to the current incomplete coverage, the points lie slightly above those of the mock catalogue. The window function is still sharply peaked in ${\mathbf k}$-space, however, and so by considering only those regions with  $k_{\parallel}$ \& $|\mathbf{k}_{\perp}|$ $ \ge0.02\;h\;$Mpc$^{-1}$, any window function effects should still be negligible.

\subsection{The effect of redshift determination errors}\label{rederr}

There is an uncertainty in determining QSO redshifts from low S/N spectra of $\delta z \sim 0.0035$ (Croom et al. 2001b), which, at the average redshift of the survey, assuming $\Lambda$CDM, corresponds to an uncertainty in the line-of-sight distance of $\sim 5\;h^{-1}\;$Mpc. This introduces an apparent velocity dispersion of $\sigma_p \sim 600\;$km$\;$s$^{-1}$. By adding this in quadrature to the intrinsic small-scale velocity dispersion in the QSO population, we therefore expect an observed velocity dispersion of $\sigma_p \sim 600 - 800\;$km$\;$s$^{-1}$ (assuming an intrinsic velocity dispersion of $\sigma_p \sim 0 - 500\;$km$\;$s$^{-1}$). A second possible source of redshift determination error, due to the intrinsic variation in line centroids between QSOs is harder to estimate. Although the intrinsic pairwise velocity dispersion is perhaps best modelled with an exponential, the apparent velocity dispersion introduced by redshift determination errors is more likely to have a Gaussian distribution, and hence ideally we should adopt a model that is a convolution of both. At the scales we are considering, however, there is virtually no difference between the two models (Ballinger et al. 1996), and so we have continued to adopt an exponential model. 

To test this we introduced a $5\;h^{-1}\;$Mpc Gaussian dispersion to the line-of-sight distances of the mock QSOs from the {\it Hubble Volume} simulation and repeated the analysis. Still assuming $\sigma_p=400\;$km$\;$s$^{-1}$, the best-fitting value of $\Omega_{\rm m}$ was systematically higher for any given value of $\beta$. As shown in section~\ref{reddist} an under-estimation of $\sigma_p$ would lead to such a systematic shift. This was removed by increasing the value of $\sigma_p$ in the model to take into account the apparent velocity dispersion introduced due to redshift determination error, assuming instead a value of  $\sigma_p=700\;$km$\;$s$^{-1}$.

As discussed in section~\ref{reddist}, and shown in figure~\ref{fig3}, $\sigma_p$ is not well constrained in this analysis, and so we have chosen to fix it at $\sigma_p=700\;$km$\;$s$^{-1}$ for the 2QZ analysis. A complementary approach, using the two-point correlation function, $\xi(\sigma,\pi)$ is more sensitive to non-linear effects, such as the velocity dispersion, as it probes smaller scales, and so should provide a stronger constraint on $\sigma_p$. A careful analysis of the errors in redshift determination of 2QZ QSOs using \textsc{autoz} would also constrain our assumed $\sigma_p$ further.

\subsection{$\chi^2$ fitting}\label{sec5}

\begin{figure}
\centerline{\hbox{\psfig{figure=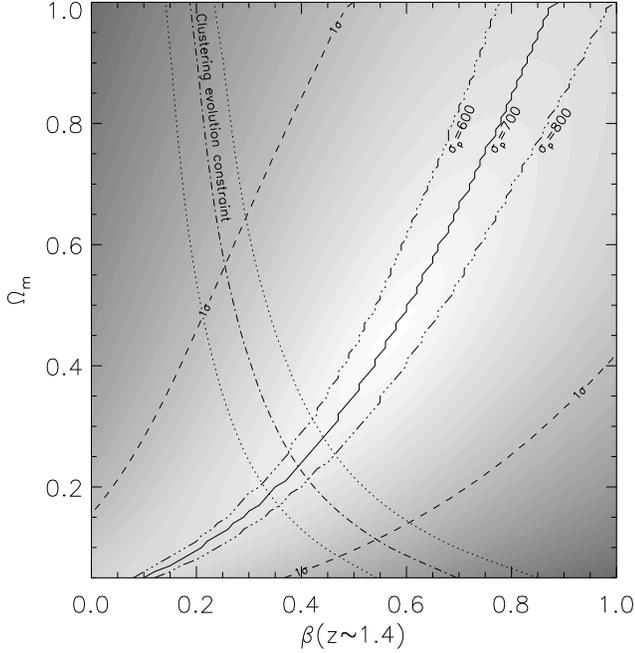,width=9.5cm}}}
\caption{Filled contours of increasing $\chi^2$ in the $\Omega_{\rm m}$ -- $\beta$ plane for fits to the 2QZ 10k catalogue. The solid line represents the best fit value of $\beta$ for each $\Omega_{\rm m}$ and the 1-$\sigma$ statistical error is given by the dashed line. The dot-dot-dot-dash lines above and below the solid line show the  best fit value of $\beta$ for each $\Omega_{\rm m}$ with $\sigma_p=600\;$km$\;$s$^{-1}$ and $\sigma_p=800\;$km$\;$s$^{-1}$ respectively.
Overlaid is the best-fit (dot-dash) and 1-$\sigma$ (dot) values of  $\beta$ determined using the mass clustering evolution method described in section~\ref{massclu}
The joint best fit values obtained are  $\beta = 0.39$ and $\Omega_{\rm m} = 0.23$.
}
\label{fig10000fit}
\end{figure}

\begin{figure}
\centerline{\hbox{\psfig{figure=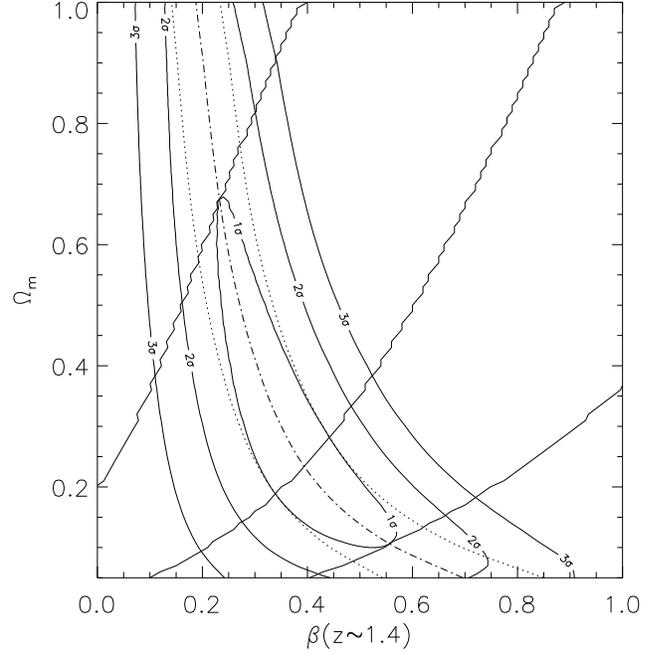,width=9.5cm}}}
\caption{Significance contours given by joint consideration of the two constraints from the 2QZ 10k catalogue. The errors from the power spectral analysis include statistical errors and the systematic error due to a $\pm100\;$km$\;$s$^{-1}$ uncertainty in $\sigma_p$. The joint best fit values obtained are  $\beta = 0.39$ and $\Omega_{\rm m} = 0.23$.}
\label{figjoint10}
\end{figure}

As before, we reconstructed a real-space power spectrum self-consistently from the measured 2QZ QSO redshift-space power spectrum (Hoyle et al. 2001a), by inverting the above redshift distortion equations for an average value of $\mu$.
The result of fitting the redshift-space distortions model to the observed power spectrum, $P^S(k_{\parallel},\mathbf{k}_{\perp})$, is shown in figure~\ref{fig10000fit}. A nominal best fit of $\Omega_{\rm m} = 0.46$, and $\beta = 0.58$ is obtained, with $\chi^2=30.74$ over 31 degrees of freedom. The fit is highly degenerate, however, and little constraint on either parameter alone can be reached. When we jointly consider the $\Omega_{\rm m}$, $\beta$ constraint from mass clustering evolution, as discussed in section~\ref{massclu}, a $\Lambda$CDM cosmology is marginally favoured with joint best fit values of $\beta = 0.39^{+0.18}_{-0.17}$ and $\Omega_{\rm m} = 0.23^{+0.44}_{-0.13}$ being obtained (see figure~\ref{figjoint10}).  The quoted errors include statistical errors and the systematic error due to a $\pm100\;$km$\;$s$^{-1}$ uncertainty in $\sigma_p$. If the error in determining QSO redshifts were over-estimated by Croom et al. (2001b) then we would find slightly lower values for $\Lambda$ and $\beta$.

This result is in line with the recent determination of $\Lambda$ from high reshift supernovae. Perlmutter et al. (1999) obtained an estimate of $\Omega_{\rm m}=0.28^{+0.09}_{-0.08}$, assuming a flat cosmology. Whilst statistically stronger than the geometric method, SN1a physics is still poorly understood and it is possible that evolutionary effects could introduce large systematic errors (Drell et al. 2000). The method described in this paper provides an independent determination of $\Lambda$ that relies only on the assumption of isotropy, and relatively well understood redshift distortion models. 

The value of $\beta_q(z\sim1.4)$ obtained is almost identical to that of local galaxies; e.g. $\beta_{\rm g}(z\sim0) = 0.43\pm0.07$ determined using the 2dF Galaxy Redshift Survey (Peacock et al. 2001). If QSOs trace average galaxy environments, as suggested by deep imaging around QSOs (e.g. Croom \& Shanks 1999), measurements of QSO clustering (Croom et al. 2001a, Hoyle et al. 2001a), and the observation that most, if not all, major galaxies appear to host a supermassive black hole (Magorrian 1998), then $\beta$ appears to vary very little as a function of redshift. With a larger QSO sample, we should be able to investigate the evolution of $\beta$ by measuring it in narrower redshift bins. The main caveat would be whether we really are comparing the same population of objects at different redshifts.

\subsection{Comparison with {\it Hubble Volume} simulation}

Using the {\it Hubble Volume} simulation, we have demonstrated that a significant constraint on the $\Lambda$ and $\beta$ parameters will be available once the QSO survey is complete. The current survey, however, has fewer QSOs and a complicated window function. To investigate if this window function has any systematic effect on the 10k results we can again turn to the {\it Hubble Volume}. 

\begin{figure}
\centerline{\hbox{\psfig{figure=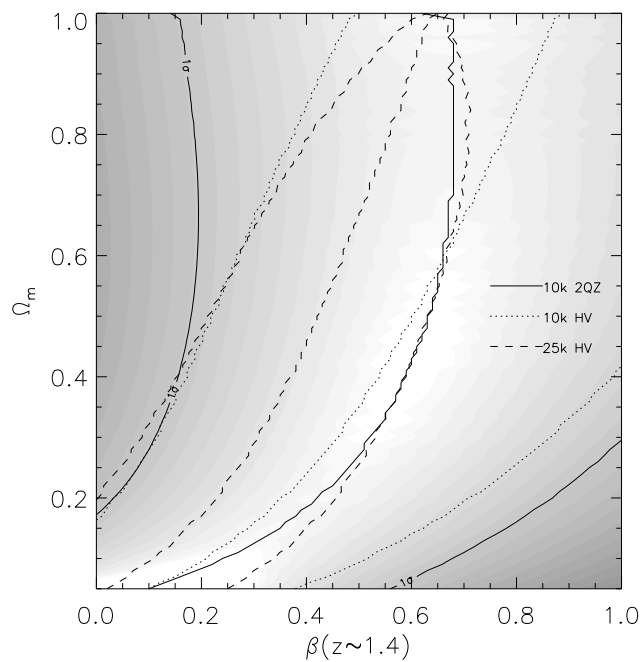,width=9.5cm}}}
\caption{Filled contours of increasing $\chi^2$ in the $\Omega_{\rm m}$ -- $\beta$ plane for fits to the {\it Hubble Volume} mock 10k 2QZ catalogue. The solid lines represent the best fit value of $\beta$ for each $\Omega_{\rm m}$ and the 1-$\sigma$ statistical errors. 
Overlaid are the best-fit and 1-$\sigma$ (dot) values determined for the 2QZ 10k catalogue (Fig.~\ref{fig10000fit}), and the best-fit and 1-$\sigma$ (dashed) values for the 25k mock catalogue (Fig.~\ref{fig3}). Taking into account the mass clustering evolution method described in section~\ref{massclu},
the joint best fit values obtained are  $\beta = 0.42$ and $\Omega_{\rm m} = 0.20$, compared  to $\beta = 0.33$ and $\Omega_{\rm m} = 0.33$ for the 25k mock survey.
}
\label{fig10000hv}
\end{figure}

We have repeated the power spectral analysis of the {\it Hubble Volume} simulation, using the current 2QZ window function when constructing a mock catalogue of 10000 QSOs. The results are shown in figure~\ref{fig10000hv}. The constraint on the $\Lambda$ and $\beta$ parameters obtained from the 10k {\it Hubble Volume} simulation is very similar to that obtained from the 10k 2QZ catalogue, although with slightly larger uncertainty. The joint best fit values obtained, taking into account the mass clustering evolution method described in section~\ref{massclu},
 are $\beta = 0.42$ and $\Omega_{\rm m} = 0.20$. Comparing the results derived from the 10k and 25k mock catalogues, there is a shift in the best fitting parameters to a slightly higher $\beta$ for a given $\Omega_{\rm m}$ when the incomplete window function is considered, however the two constraints are entirely consistent within the errors, indicating that possible systematic effects due to the 10k window function are fairly small. If this shift is systematic then taking it into account would slightly weaken the significance of our $\Lambda$ constraint from the 10k 2QZ catalogue.

\section{Summary and Conclusions}

When the 2QZ is complete, a powerful geometric test for the
cosmological constant (Alcock \& Paczy\'{n}ski 1979) will be available.
The distance between a pair of QSOs measured along the line of sight,
calculated from the relative redshifts, has a different dependence  on
the cosmological parameters than the distance measured in the angular
direction.  By comparing the clustering along and across the line of
sight and modelling the effects of peculiar velocities and bulk
motions in redshift space, geometric and redshift-space distortions can be detected.  

In this paper we have examined the potential of this test by considering the anisotropy in the 2QZ power spectrum. Ballinger et al. (1996) developed a model to estimate $\Lambda$ and $\beta$ from the geometric and redshift-space distortions seen in the power spectra of redshift surveys. We have applied this model to a detailed simulation of the 2QZ, produced using the Virgo Consortium's huge {\it Hubble Volume} N-body $\Lambda$-CDM light cone simulation. The results confirm the conclusions of Ballinger et al.; the shape of the redshift-space and geometric distortions are very similar, and discriminating between the two to produce a purely geometric constraint on $\Lambda$ is difficult. When all the uncertainties in measuring $P^S(k_{\parallel},\mathbf{k}_{\perp})$ for the 2QZ are taken into account we have found that only a joint $\Lambda - \beta$ constraint is possible. 

However, by combining this result with a second constraint based on mass clustering evolution, we can make significant progress. We predict that this should allow us to constrain $\beta$ to approximately $\pm0.1$, and $\Omega_{\rm m}$ to $\pm0.25$ in the final 25k survey. Note that throughout this paper we have assumed a flat cosmology. 

One possible caveat with this technique is that we are comparing values of $\beta_g$ and $\beta_q$ that were potentially measured using different estimators and on different scales. To minimise any possible systematic effects the value of $\beta_g$ could be measured using the same method (Outram, Hoyle \& Shanks 2000). Although at a much lower redshift, there will also still be a small cosmological dependence in the result which should be taken into account.

Applying this analysis to the 2QZ 10k Catalogue we find that a $\Lambda$CDM cosmology is favoured with joint best fit values of $\beta = 0.39^{+0.18}_{-0.17}$ and $\Omega_{\rm m} = 0.23^{+0.44}_{-0.13}$ being obtained. An uncertainty in determining QSO redshifts from low S/N spectra of $\delta z \sim 0.0035$ introduces an apparent velocity dispersion of $\sigma_p \sim 600\;$km$\;$s$^{-1}$.  The quoted errors include a systematic error due to a $\pm100\;$km$\;$s$^{-1}$ uncertainty in $\sigma_p$. A careful analysis of the errors in redshift determination of 2QZ QSOs using \textsc{autoz} would help constrain this; if the errors in determining QSO redshifts are currently over-estimated then we would find slightly lower values for $\Lambda$ and $\beta$. A second possible source of redshift determination error, due to the intrinsic variation in line centroids between QSOs, is harder to estimate.

A complementary approach, using the two-point correlation function, $\xi(\sigma,\pi)$, to investigate redshift-space distortions in QSO clustering can also be considered (Matsubara \& Suto 1996). This approach probes smaller scales, and so is more sensitive to non-linear effects, such as the velocity dispersion. 
A second paper will investigate the application of this analysis to the 2QZ (Hoyle et al. 2001a).

\section*{Acknowledgements}

\noindent The 2QZ is based on observations made with the Anglo-Australian Telescope and the UK Schmidt Telescope, and we would like to thank our colleagues on the 2dF galaxy redshift survey team and all the staff at the AAT that have helped to make this survey possible. We would like to thank Adrian Jenkins, Gus Evrard, and the Virgo Consortium for kindly providing the {\it Hubble Volume} simulations. FH \& NL acknowledge the receipt of a PPARC studentship.

\end{document}